\documentclass[seceq,preprint]{ptptex}
\usepackage{wrapft}
\usepackage{graphicx}

\newcommand{\rbra}[1]{( {#1} |}     
\newcommand{\rket}[1]{| {#1} )}     
\newcommand{\wtilde}[1]{\widetilde{#1}} 

\newcommand{\kbar}{k \kern -0.5em\raise 0.6ex \hbox{--}}

\def\<{\langle}
\def\>{\rangle}
\def\bsub{\begin{subequations}}
\def\esub{\end{subequations}}
\def\beqn{\begin{eqnarray}}
\def\eeqn{\end{eqnarray}}
\def\b{\begin{equation}}

\title{
Semi-Classical Approach to the Two-Level Pairing Model
}
\subtitle{
Various Aspects of Phase Change
}    

\author{
Yasuhiko {\sc Tsue},$^{1}$ 
Constan\c{c}a {\sc Provid\^encia},$^{2}$\\
Jo\~ao da {\sc Provid\^encia}$^{2}$
and Masatoshi {\sc Yamamura}$^{3}$
}

\inst{
$^1$Physics Division, Faculty of Science, Kochi University, Kochi 780-8520, 
Japan\\
$^{2}$Departamento de F\'isica, Universidade de Coimbra, 3004-516 Coimbra, 
Portugal\\
$^{3}$Faculty of Engineering, Kansai University, Suita 564-8680, Japan
}

\recdate{
\today
}

\abst{
Aspects of the phase change of the two-level pairing model are investigated 
in the semi-classical treatment by using the variational 
approch with the mixed-mode coherent state. 
In the classical limit, $\hbar\rightarrow 0$, the sharp phase transition 
appears and the two phases exist in the region where the force strength 
is larger than a certain critical value. 
However, it is shown that, in the semi-classical treatment, 
the above-mentioned behavior of the phase change disappears in the analytical 
and the numerical treatments. 
As a result, new understanding about aspects of the phase change in this model 
is given. 
}
\begin{document}
\maketitle

\section{Introduction}

The phase structure and the phase transitions 
between various phases in condensed matter systems, such as nuclear 
and/or quark matter as well as the condensed matter in atoms and molecules, 
are important concepts to understand the nature of the matter 
under consideration. 
In the infinite systems, the clear phase separation is often realized and, 
in that case, the sharp phase transition may occur among the different phases. 
On the other hand, in finite systems such as nucleus, the phase change is 
not so clear. This situation can be demonstrated\cite{3} 
in the exactly-solvable many-particle 
models such as the Lipkin model\cite{1} or the pairing model.\cite{2} 
However, the mean field approximation in the time-independent theory, 
which often leads to the classical treatment, gives the fictitious sharp phase 
transition. 
Addition to the above mentioned problem, the two phases, namely, the 
condensed and the non-condensed phases, appear in both the Lipkin and 
the pairing models in the classical treatment, 
if the force strength is larger than a certain critical value at zero 
temperature.

In this paper, an exactly-solvable two-level pairing model is 
treated to understand the nature of the phase change and the existence 
of two phases for the force strength at zero temperature 
in the classical limit. 
The two-level pairing model is regarded as the $su(2)\otimes su(2)$-algebraic 
model. 
For the original many-fermion system, the trial state for the variation 
is restricted to the BCS state which gives the mean field approximation 
and the unique classical counterpart for the original quantal system. 
In order to avoid the usual mean field approximation of the BCS theory, 
the two-level pairing model can be expressed in the four kinds of boson 
operators by using the Schwinger boson representation of the $su(2)$-algebra. 
In this boson representation, the various trial states are 
possible to obtain a classical counterpart.\cite{4} 
Thus, in this paper, a certain wave packet which we call the mixed-mode 
coherent state\cite{A} is adopted to the variational calculation. 
This state gives the semi-classical approximation which retains $\hbar$

This paper is organized as follows: 
In the next section, the two-level pairing model is introduced by means of 
the Schwinger boson representation. In \S 3, the classical counterpart 
for the two-level pairing model is given by using the mixed-mode coherent 
state and the basic equations of motion in this approach are 
also derived. 
In \S 4, the energy minimum point is investigated in the classical limit 
$\hbar\rightarrow 0$ and in the semi-classical approach with finite value 
of $\hbar$. 
Then, it is shown that the sharp phase transition does not occur with 
finite $\hbar$, while 
the sharp phase transition occurs and two phases exist in the certain 
region in the classical limit. 
The method to obtain an approximate solution for the energy minimum 
is given analytically in \S 5 together with appendix A. 
The behavior of the phase change is explained with unknown concept in \S 6 
and 
the discussion about the ground state energy is given in \S 7. 
The last section is devoted to a summary.

\section{Two-level pairing model in many-fermion system and its counterpart 
in the Schwinger boson representation}

The model discussed in this paper is two-level pairing model 
in many-fermion system. 
The two levels are specified by $\sigma=+$ (the upper) and 
$\sigma=-$ (the lower), respectively. 
The difference of the single-particle energies between the two levels 
and the strength of the pairing interaction are denoted as 
$\hbar \epsilon \ (>0)$ and $\hbar^2 G\ (>0)$, respectively. 
The Hamiltonian, ${\wtilde H}$, is expressed in the form 
\begin{equation}\label{2-1}
{\wtilde H}=\epsilon({\wtilde S}_0(+)-{\wtilde S}_0(-))
-G({\wtilde S}_+(+)+{\wtilde S}_+(-))
({\wtilde S}_-(+)+{\wtilde S}_-(-))  \ .
\end{equation}
Here, ${\wtilde S}_0(\sigma)$, ${\wtilde S}_+(\sigma)$ and 
${\wtilde S}_-(\sigma)$ are defined as 
\begin{subequations}\label{2-2}
\begin{eqnarray}
& &{\wtilde S}_0(\sigma)=(\hbar/2)({\wtilde N}(\sigma)-\Omega_{\sigma}) \ , 
\label{2-2a}\\
& &{\wtilde S}_+(\sigma)=(\hbar/2){\wtilde P}^*(\sigma) \ , \qquad
{\wtilde S}_-(\sigma)=(\hbar/2){\wtilde P}(\sigma) \ , 
\qquad\qquad\qquad\qquad\qquad\qquad\qquad\ \ 
\label{2-2b}
\eeqn
\end{subequations}
\vspace{-0.9cm}
\begin{subequations}\label{2-3}
\beqn
& &{\wtilde N}(\sigma)=\sum_{m=-j_\sigma}^{j_{\sigma}}{\tilde c}_m^*(\sigma)
{\tilde c}_m(\sigma) \ , \qquad \Omega(\sigma)=(2j_\sigma+1)/2\ , 
\label{2-3a}\\ 
& &{\wtilde P}^*(\sigma)=\sum_{m={-j_\sigma}}^{j_{\sigma}}
{\tilde c}_m^*(\sigma)(-)^{j_\sigma-m}{\tilde c}_{-m}^*(\sigma)\ , \ 
{\wtilde P}(\sigma)=\sum_{m={-j_\sigma}}^{j_{\sigma}}(-)^{j_\sigma-m}
{\tilde c}_{-m}(\sigma){\tilde c}_{m}(\sigma)\ .\qquad
\label{2-3b}
\end{eqnarray}
\end{subequations}
The operator (${\tilde c}_m^*(\sigma),{\tilde c}_m(\sigma))$ denotes 
fermion operator in the single-particle state 
$(\sigma, j_\sigma,m\ ; m=-j_\sigma, -j_\sigma+1,\cdots ,
j_\sigma-1,j_\sigma)$. 
Of course, $j_\sigma$ is a half-integer. Clearly, ${\wtilde N}(\sigma)$ 
and $({\wtilde P}^*(\sigma), {\wtilde P}(\sigma))$ denote the fermion 
number and the fermion pair operators in the state $\sigma$, respectively.

We know that the set $({\wtilde S}_{\pm,0}(\sigma))$ obeys the 
$su(2)$-algebra, and then, essentially, the presented model is governed by 
the $su(2)\otimes su(2)$-algebra. 
Clearly, the addition of the two sets also obeys the $su(2)$-algebra: 
\begin{equation}\label{2-4}
{\wtilde S}_{\pm,0}={\wtilde S}_{\pm,0}(+)+{\wtilde S}_{\pm,0}(-) \ .
\end{equation}
The Hamiltonian (\ref{2-1}) has three constants of motion. 
This can be shown from the following relation: 
\begin{equation}\label{2-5}
[\ {\wtilde {\mib S}}^2(+)\ , \ {\wtilde H}\ ]=
[\ {\wtilde {\mib S}}^2(-)\ , \ {\wtilde H}\ ]=
[\ {\wtilde S}_0\ , \ {\wtilde H}\ ]=0 \ .
\end{equation}
Here, ${\wtilde {\mib S}}^2(\sigma)$ denotes the Casimir operator for 
$({\wtilde S}_{\pm,0}(\sigma))$. 
It should be noted that the Casimir operators for $({\wtilde S}_{\pm,0})$, 
${\wtilde {\mib S}}^2$, does not commute with ${\wtilde H}$: 
\begin{equation}\label{2-6}
[\ {\wtilde {\mib S}}^2\ , \ {\wtilde H}\ ] \neq 0 \ . 
\end{equation}

A general framework of the $su(2)\otimes su(2)$-algebra can be formulated in a
form of the Schwinger boson representation, in which four kinds of boson 
operators are used. 
A possible example of this formulation was performed by the present 
authors (J.P. and M.Y.) with Kuriyama.\cite{A} 
Hereafter, we will refer it as (A). 
In (A), the $su(2)\otimes su(2)$-algebra is formulated in terms of the 
bosons $({\hat a}_\sigma^*, {\hat a}_{\sigma}, {\hat b}_{\sigma}^* , 
{\hat b}_{\sigma} ; \sigma=\pm)$. 
In this framework, we can construct a counterpart of the model 
presented in this section in the Schwinger boson representation. 
The $su(2)$-generators $({\wtilde S}_{\pm,0}(\sigma);\sigma=\pm)$ 
correspond to ${\hat S}_{\pm,0}(\sigma)$ in the form 
\beqn
& &{\wtilde S}_{\pm,0}(\sigma) \longrightarrow {\hat S}_{\pm,0}(\sigma) \ , 
\label{2-7}\\
& &{\hat S}_0(\sigma)=(\hbar/2)({\hat a}_{\sigma}^*{\hat a}_{\sigma}
-{\hat b}_{\sigma}^*{\hat b}_{\sigma}) \ , \nonumber\\
& &{\hat S}_+({\sigma})=\hbar{\hat a}_{\sigma}^*{\hat b}_{\sigma} \ , 
\qquad
{\hat S}_-(\sigma)=\hbar{\hat b}_{\sigma}^*{\hat a}_{\sigma}\ . 
\label{2-8}
\eeqn
It should be noted that the notations are different from those in (A). 
Of course, we have 
\beqn\label{2-9}
& &{\hat S}_0=(\hbar/2)({\hat a}_+^*{\hat a}_+ - {\hat b}_+^*{\hat b}_+
+{\hat a}_-^*{\hat a}_- - {\hat b}_-^*{\hat b}_-) \ , \nonumber\\
& &{\hat S}_+=\hbar({\hat a}_+^*{\hat b}_+ + {\hat a}_-^*{\hat b}_-) \ , 
\nonumber\\
& &{\hat S}_-=\hbar({\hat b}_+^*{\hat a}_+ + {\hat b}_-^*{\hat a}_-) \ .
\eeqn
Associating with the above operators, we can define the following 
operators: 
\begin{equation}\label{2-10}
{\hat S}(\sigma)=(\hbar/2)({\hat a}_{\sigma}^*{\hat a}_{\sigma}
+{\hat b}_{\sigma}^*{\hat b}_{\sigma}) \ . 
\end{equation}
This operator satisfies, for the Casimir operator 
${\hat {\mib S}}^2(\sigma)$, the relation 
\begin{equation}\label{2-11}
{\hat {\mib S}}^2(\sigma)={\hat S}(\sigma)({\hat S}(\sigma)+\hbar) \ .
\end{equation}
In the case where the seniorities are equal to 0 in the fermion system, 
we should set up the correspondence 
\begin{equation}\label{2-12}
(\hbar/2)\Omega(\sigma) \longrightarrow {\hat S}(\sigma) \ .
\end{equation}
In the Schwinger boson representation, there does not exist 
the concept of the fermion number explicitly. 
The definition of ${\wtilde S}_0(\sigma)$ given in the relation 
(\ref{2-2a}) and the correspondence (\ref{2-12}) permit us to 
introduce the fermion number in the Schwinger boson 
representation in the form 
\begin{equation}\label{2-13}
\hbar{\hat N}(\sigma)=2({\hat S}(\sigma)+{\hat S}_0(\sigma))
=2\hbar {\hat a}_{\sigma}^*{\hat a}_{\sigma} \ .
\end{equation}
Therefore, the total number ${\hat N}$ is given as 
\beqn\label{2-14}
\hbar{\hat N}=\hbar({\hat N}(+)+{\hat N}(-))&=&
2({\hat S}(+)+{\hat S}(-)+{\hat S}_0) \nonumber\\
&=&2\hbar ({\hat a}_+^*{\hat a}_+ + {\hat a}_-^*{\hat a}_-) \ .
\eeqn
From the above form, we can see that in the present case for both levels, 
even fermion numbers should be occupied. 
In the case of odd fermion numbers, we must take into account 
the case where the seniorities do not vanish.

\section{Classical counterpart of the $su(2)\otimes su(2)$-algebra}

We continue the recapitulation of (A). In (A), we showed that the 
orthogonal set for the $su(2)\otimes su(2)$-algebra is obtained with the 
use of the set $({\hat R}_{\pm,0},{\hat T}_{\pm,0})$ as a supporting role. 
The set is defined in the form 
\beqn
& &{\hat R}_0=(\hbar/2)({\hat a}_+^*{\hat a}_+
-{\hat a}_-^*{\hat a}_-) \ , \nonumber\\
& &{\hat R}_+=\hbar({\hat a}_+^*{\hat a}_- +{\hat b}_+^*{\hat b}_-) \ , \qquad
{\hat R}_-=\hbar({\hat a}_-^*{\hat a}_+ +{\hat b}_-^*{\hat b}_+) \ , 
\label{3-1}\\
& &{\hat T}_0=(\hbar/2)({\hat a}_+^*{\hat a}_+ +{\hat b}_-^*{\hat b}_-
+{\hat a}_-^*{\hat a}_- +{\hat b}_+^*{\hat b}_+ +2) \ , \nonumber\\
& &{\hat T}_+=\hbar({\hat a}_+^*{\hat b}_-^* -{\hat a}_-^*{\hat b}_+^*) \ , 
\qquad
{\hat T}_-=\hbar({\hat b}_-{\hat a}_+ -{\hat b}_+{\hat a}_-) \ , 
\label{3-2}
\eeqn
The sets $({\hat R}_{\pm,0})$ and $({\hat T}_{\pm,0})$ obey the 
$su(2)$- and the $su(1,1)$-algebra, respectively, and it is characteristic 
that three sets $({\hat S}_{\pm,0})$, $({\hat R}_{\pm,0})$ and 
$({\hat T}_{\pm,0})$ commute with one another. 
Further, the Casimir operators for the three forms are identically 
equal to one another. 

In (A), we defined a wave packet, which we called a mixed-mode coherent 
state, as follows: 
\beqn\label{3-3}
\rket{c}&=&
U_t^{-2}\exp \left(-|W|^2/\hbar\right)\cdot
\exp\left((V_t/\hbar U_t){\hat T}_+\right)\cdot
\exp\left((V_s/\hbar U_s){\hat S}_+\right)
\nonumber\\
& &\times \exp\left((V_r/\hbar U_r){\hat R}_+\right)\cdot
\exp\left(\left(\sqrt{\hbar/2}WU_r U_s/U_t\right){\hat b}_-^*\right)
\rket{0} \ . 
\eeqn
The state $\rket{c}$ is normalized and $U_r$, $U_s$ and $U_t$ are real 
and $W$, $V_r$, $V_s$ and $V_t$ complex. 
They obey the condition 
\begin{equation}\label{3-4-0}
U_t^2-|V_t|^2=1\ , \quad
U_s^2+|V_s|^2=1\ , \quad
U_r^2+|V_r|^2=1\ . 
\end{equation}
We can prove the following relation: 
\beqn\label{3-4}
& &{\hat a}_+=U_t{\hat A}_+ + V_t{\hat B}_-^*
+\sqrt{\frac{2}{\hbar}}(WV_r V_s U_t + W^*U_r U_s V_t) \ , \nonumber\\
& &{\hat a}_-=U_t{\hat A}_- - V_t{\hat B}_+^*
+\sqrt{\frac{2}{\hbar}}(WU_r V_s U_t - W^*V_r^* U_s V_t) \ , \nonumber\\
& &{\hat b}_+=U_t{\hat B}_+ - V_t{\hat A}_-^*
+\sqrt{\frac{2}{\hbar}}(WV_r U_s U_t - W^*U_r V_s^* V_t) \ , \nonumber\\
& &{\hat b}_-=U_t{\hat B}_- + V_t{\hat A}_+^*
+\sqrt{\frac{2}{\hbar}}(WU_r U_s U_t + W^*V_r^* V_s^* V_t) \ .
\eeqn
Here, $({\hat A}_\sigma , {\hat A}_{\sigma}^*)$ and 
$({\hat B}_{\sigma}, {\hat B}_{\sigma}^*)$ denote boson operators satisfying 
\begin{equation}\label{3-5}
{\hat A}_{\sigma}\rket{c}=0 \ , \qquad {\hat B}_{\sigma}\rket{c}=0 \ .
\end{equation}
As a possible parameterization, in (A), the following form 
were presented: 
\beqn\label{3-6}
& &W=\sqrt{S}e^{-i(\psi-\psi_0+\psi(-))/2} \ , \nonumber\\
& &V_r=\sqrt{(S+S(+)-S(-))/2S}e^{-i(\psi(+)-\psi(-))/2} \ , \nonumber\\
& &V_s=\sqrt{(S+S_0)/2S}e^{-i\psi_0} \ , \nonumber\\
& &V_t=\sqrt{(S(+)+S(-)-S)/2(S+\hbar)}e^{-i(\psi(+)+\psi(-))/2} \ , \nonumber\\
& &U_r=\sqrt{(S-S(+)+S(-))/2S} \ , \qquad
U_s=\sqrt{(S-S_0)/2S} \ , \nonumber\\
& &U_t=\sqrt{(S(+)+S(-)+S+2\hbar)/2(S+\hbar)} \ .
\eeqn
In (A), we prove that $(\psi(\sigma),S(\sigma))$, $(\psi_0, S_0)$ and 
$(\psi, S)$ are in the relations of the angle and the action variables 
in classical mechanics. 
Further, we have the following relation: 
\begin{equation}\label{3-7}
S \geq 0 \ , \qquad
-S \leq S_0 \leq S \ , \qquad
|S(+)-S(-)| \leq S \leq S(+)+S(-) \ . 
\end{equation}
The above is nothing but the relation that implies the coupling of 
two $su(2)$-spins.

With the aid of the relation (\ref{3-4}), we can calculate the expectation 
value of any operator composed of the present bosons with respect to 
$\rket{c}$: 
We denote $\rbra{c}{\hat O}\rket{c}=({\hat O})_c$. 
For example, we have 
\beqn
& &({\hat S}(\sigma))_c=S(\sigma) \ , 
\label{3-8}\\
& &({\hat S}_0)_c=S_0 \ , \nonumber\\
& &({\hat S}_{\pm})_c=\sqrt{S^2-S_0^2}\ e^{\pm i\psi_0} \ .
\label{3-9}
\eeqn
The relation (\ref{3-9}) is identical with the classical counterpart of 
the Holstein-Primakoff boson representation, and then, including the 
relation (\ref{3-7}), our present formalism may be called 
the classical counterpart of the $su(2)\otimes su(2)$-algebra. 
Concerning ${\hat S}_+{\hat S}_-$, the most interesting point is as 
following: 
\beqn
& &({\hat S}_+{\hat S}_-)_c
=(({\hat S}_+(+)+{\hat S}_+(-))({\hat S}_-(+)+{\hat S}_-(-)))_c
=S^2-S_0^2+\hbar(S+S_0) \ . \quad
\label{3-10}\\
& &({\hat S}_+)_c({\hat S}_-)_c=S^2-S_0^2 \ . 
\label{3-11}
\eeqn
We can see that in the relation (\ref{3-10}), the quantal fluctuation is 
exactly taken into account. 
Further, we have 
\beqn\label{3-12}
& &({\hat S}_0(+)-{\hat S}_0(-))_c \nonumber\\
&=&\frac{(S(\!+\!)-S(\!-\!))(S(\!+\!)+S(\!-\!)+\hbar)S_0}
{S(S+\hbar)}\nonumber\\
& &+\frac{\sqrt{(S\!-\!S(\!+\!)\!+\!S(\!-\!))(S\!+\!S(\!+\!)\!-\!S(\!-\!))
(S(\!+\!)\!+\!S(\!-\!)\!-\!S)(S(\!+\!)\!+\!S(\!-\!)\!+\!S\!+\!2\hbar)}
}{S(S+\hbar)}\cos \psi \ . \nonumber\\
& &
\eeqn

Under the above preparation, we are able to obtain the classical Hamiltonian 
of ${\hat H}$ which is the counterpart of ${\wtilde H}$ shown in the 
relation (\ref{2-1}):
\begin{equation}\label{3-13}
{\hat H}=\epsilon({\hat S}_0(+)-{\hat S}_0(-))
-G({\hat S}_+(+)+{\hat S}_+(-))({\hat S}_-(+)+{\hat S}_-(-)) \ .
\end{equation}
The Hamiltonian ${\hat H}$ obeys the condition 
\begin{equation}\label{3-14}
[\ {\hat S}(+)\ , \ {\hat H}\ ]=
[\ {\hat S}(-)\ , \ {\hat H}\ ]=
[\ {\hat S}_0\ , \ {\hat H}\ ]=0 \ .
\end{equation}
The relation (\ref{3-14}) corresponds to the relation (\ref{2-5}). 
It is enough to calculate the expectation value of ${\hat H}$ for $\rket{c}$: 
$({\hat H})_c=H$. 
In this case, the forms (\ref{3-10}) and (\ref{3-12}) are useful. 
Of course, the classical expression of the fermion number is given in the form 
\begin{equation}\label{3-16}
\hbar N=\hbar({\hat N})_c=2(S(+)+S(-)+S_0) \ . 
\end{equation}
For the above, the relation (\ref{2-14}) is used. 
Corresponding to the Hamiltonian ${\hat H}$, $H$ satisfies 
\begin{equation}\label{3-17}
[\ {S}(+)\ , \ {H}\ ]_P=
[\ {S}(-)\ , \ {H}\ ]_P=
[\ {S}_0\ , \ {H}\ ]_P=0 \ .
\end{equation}
Here, $[\ , \ ]_P$ denotes the Poisson bracket. 
As was already mentioned, the parameters $(\psi(\sigma), S(\sigma))$, 
$(\psi_0, S_0)$ and $(\psi, S)$ are in the relation of 
the canonical variables in classical mechanics. 
From the relation (\ref{3-17}), we can see that $S(+)$, $S(-)$ and $S_0$ 
are constants of motion and the time-evolution of the angles 
$\psi(+)$, $\psi(-)$ and $\psi_0$ are given in terms of 
\begin{equation}\label{3-18}
{\dot \psi}(+)=[\ \psi(+)\ , \ H\ ]_P \ , \qquad
{\dot \psi}(-)=[\ \psi(-)\ , \ H\ ]_P \ , \qquad
{\dot \psi}_0=[\ \psi_0\ , \ H\ ]_P \ .
\end{equation}
The variables $(\psi, S)$ satisfy the Hamilton equations of motion: 
\begin{equation}\label{3-19}
{\dot \psi}=[\ \psi\ , \ H\ ]_P \ , \qquad
{\dot S}=[\ S\ , \ H\ ]_P \ .
\end{equation}

In order to avoid unnecessary complication, we treat the following case: 
\beqn\label{3-20}
& &S(+)=S(-)\ (=L)\ , \qquad 
L=\frac{\hbar}{2}\Omega\ , \quad (\Omega(+)=\Omega(-)=\Omega) \nonumber\\
& &S_0=0 \ .
\eeqn
In this case, the relation (\ref{3-16}) is reduced to 
\begin{equation}\label{3-21}
\hbar N=4L \ , \qquad {\rm i.e.,}\qquad N=2\Omega \ . 
\end{equation}
The relation (\ref{3-21}) shows us that we are interested in the 
closed shell system. 
Under the expectation values (\ref{3-11}) and (\ref{3-12}) and the condition 
(\ref{3-20}), $H$ can be reduced to 
\beqn\label{3-22}
H&=&\epsilon\left(1-\frac{\hbar}{S+\hbar}\right)
\sqrt{(2L-S)(2L+S+2\hbar)}\cos \psi 
-G(S^2+\hbar S) \ .
\eeqn
The Hamiltonian (\ref{3-22}) can be rewritten as 
\beqn\label{3-23}
H&=&\epsilon\left(1-\frac{\hbar}{\sigma}\right)
\sqrt{(2\lambda)^2-\sigma^2}\cos \psi 
-G\sigma(\sigma-\hbar) \ .
\eeqn
Here, $\lambda$ and $\sigma$ are defined by 
\begin{equation}\label{3-24}
\lambda=L+\hbar/2 \ , \qquad \sigma=S+\hbar \ . 
\end{equation}
Concerning the parameters $\lambda$ and $\sigma$, the following 
relation should be noted: 
Since there exists the restriction $|S(+)-S(-)|\leq S \leq S(+)+S(-)$ 
classically and quantum mechanically, for the case $S(+)=S(-)$, we have 
\begin{equation}\label{3-25}
\hbar \leq \sigma \leq 2\lambda \ . 
\end{equation}
Since $(\psi, \sigma)$ is also in the canonical relation, we have the 
following Hamilton's equations: 
\begin{subequations}\label{3-26}
\beqn
{\dot \psi}=[\ \psi\ , \ H\ ]_P
=\frac{\partial H}{\partial \sigma}
&=&\epsilon\left[
\frac{\hbar}{\sigma^2}\sqrt{(2\lambda)^2-\sigma^2}-\left(1-\frac{\hbar}{\sigma}
\right)\frac{\sigma}{\sqrt{(2\lambda)^2-\sigma^2}}\right]\cos \psi
\nonumber\\
& &-G(2\sigma-\hbar) \ , 
\label{3-26a}\\
{\dot \sigma}=[\ \sigma\ , \ H\ ]_P
=-\frac{\partial H}{\partial \psi}
&=&
\epsilon\left(1-\frac{\hbar}{\sigma}\right)\sqrt{(2\lambda)^2-\sigma^2}
\sin\psi \ . 
\label{3-26b}
\eeqn
\end{subequations}

\section{Determination of the energy minimum point}

One of our interests in this paper 
is to investigate how the energy minimum point 
induced by the mixed-mode coherent state $\rket{c}$ changes in the phase space 
$(\psi,\sigma)$ as a function $(\epsilon, G)$. 
The energy minimum point should be stationary in the phase space, 
and then, we have the condition 
\begin{equation}\label{4-1}
{\dot \psi}=0 \ , \qquad {\dot \sigma}=0 \ . 
\end{equation}
The relation (\ref{3-26b}) gives us 
\begin{equation}\label{4-2}
\sin\psi=0 \ , \quad {\rm i.e.,}\quad \cos\psi=-1\ {\rm or}\ +1. 
\end{equation}
It may be self-evident that the form (\ref{3-23}) supports 
\begin{equation}\label{4-3}
\cos\psi=-1 \ .
\end{equation}
The angle $\psi$ does not depend on $(\epsilon, G)$. 
Under the conditions (\ref{4-1}) and (\ref{4-3}), the relation 
(\ref{3-26a}) is reduced, after some calculations, to 
\beqn\label{4-4}
& &\sigma^3\left(\epsilon^2-4G^2((2\lambda)^2-\sigma^2)\right)\nonumber\\
=& &\hbar\left(\epsilon(2\lambda)^2-G\sigma^2\sqrt{(2\lambda)^2-\sigma^2}
\right)
\left(\epsilon+2G\sqrt{(2\lambda)^2-\sigma^2}\right) \ . 
\eeqn
By solving Eq.(\ref{4-4}), we can determine the value of $\sigma$ 
which makes the energy minimum. 
As a possible solution of the latter of Eq.(\ref{4-1}), we obtain 
$\sigma=2\lambda$, which is independent of $\psi$ and $\epsilon$. 
Substituting $\sigma=2\lambda$ into the former of Eq.(\ref{4-1}), 
we have 
$\cos \psi=-(4G\lambda/\epsilon)(4\lambda-\hbar)/(4\lambda-2\hbar)\cdot
\sqrt{\lambda\delta}$. 
Here, $\delta$ denotes an infinitesimal parameter defined as 
$\sigma=2\lambda-\delta$. 
Then, under $\delta=0$, $\cos \psi=0$, i.e., $\psi=\pm \pi/2$, 
which is independent of $\epsilon$ and $G$. 
But, this solution cannot connect to the case 
$\sigma < 2\lambda$. 
Therefore, we do not adopt it. 
If we adopt the solution $\cos \psi=-1$, clearly, $4G\lambda/\epsilon$ 
becomes infinite under the condition $\delta\rightarrow 0$. 
This means that our two levels are degenerate.

Let us search the solution of Eq.(\ref{4-4}). 
This relation can be simplified in the form 
\beqn\label{4-5}
\rho^3\left(1-{x^2}(1-\rho^2)\right)
=& &\kbar\left(1-\frac{x\rho^2}{2}\sqrt{1-\rho^2}\right)
\left(1+x\sqrt{1-\rho^2}\right) \ . 
\eeqn
Here, $\kbar$, $\rho$ and $x$ are defined as 
\begin{eqnarray}
& &\kbar=\hbar/2\lambda \ , 
\label{4-6}\\
& &\rho=\sigma/2\lambda \ , \qquad 
x=4G\lambda/\epsilon\ . 
\label{4-7}
\end{eqnarray}
The relation (\ref{3-24}) and the condition $L\geq 0$ give us 
\begin{equation}\label{4-8}
0 < \kbar \leq 1 \ .
\end{equation}
Especially, we have 
\beqn\label{4-9}
& &\kbar \rightarrow 0 \ , \quad {\rm if} \quad \lambda \rightarrow \infty \ , 
\quad {\rm i.e.,}\quad L\rightarrow \infty \ , 
\nonumber\\
& &\kbar =1 \ , \quad {\rm if} \quad \lambda =\frac{\hbar}{2} \ , 
\quad {\rm i.e.,}\quad L=0 \ . 
\eeqn
If we take into account the correspondence to the original fermion system, 
the minimum value of $\Omega$ is equal to 1, and then, $L=\hbar/2$ and 
we have 
\begin{equation}\label{4-10}
0 < \kbar \leq 1/2 \ .
\end{equation}
However, formally, we have the minimum value of $L=0$. 
Further, the relation (\ref{3-25}) gives us 
\begin{equation}\label{4-11}
\kbar \leq \rho \leq 1 \ .
\end{equation}
Of course, we have 
\begin{equation}\label{4-12}
x \geq 0 \ .
\end{equation}

First of all, on the basis of the relation (\ref{4-5}), we treat the case 
$\hbar$ (the Planck constant) $=0$. 
In this case, we have a simple solution:
\begin{eqnarray}\label{4-13}
& &x=
\begin{cases} 
\hbox{\rm any\ value\ }(\geq 0)\ , \quad {\rm for}\quad \rho=0\ , 
\\
\frac{1}{\sqrt{1-\rho^2}}\ , \quad {\rm for}\quad 
0 < \rho < 1\ . \quad (x\rightarrow \infty\ , {\rm if}\ \rho\rightarrow 1)
\end{cases}
\end{eqnarray}
Inversely, the solution (\ref{4-13}) are given as 
\begin{eqnarray}\label{4-15}
& &\rho=
\begin{cases} 
0 \ , \quad \hbox{\rm for\ any\ value\ of\ }x (\geq 0)\ , \\
\frac{\sqrt{x^2-1}}{x}\ , \quad {\rm for}\quad 
x \geq 1 \ . 
\end{cases}
\end{eqnarray}
\begin{figure}[t]
\begin{center}
\includegraphics[height=7.0cm]{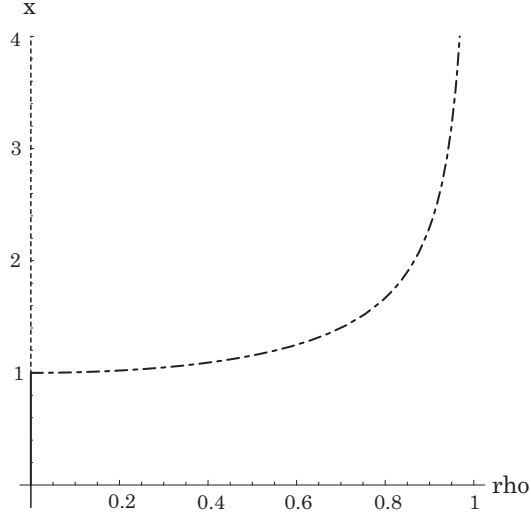}
\caption{The variable $x$ is depicted as a function of 
$\rho$ with $\hbar=0$.
}
\label{fig:4-1}
\end{center}
\end{figure}
In Fig.{\ref{fig:4-1}}, 
we see a straight line consisting of solid and dotted lines 
along $x$-axis. 
This is one phase which is conventionally called as normal 
conducting phase. 
Further, we see that at the point $x=1$ on the straight line 
($x$-axis), a new 
phase is born toward the rectangular direction and it grows up 
to a branch, which is conventionally called as superconducting phase. 
In Fig.{\ref{fig:4-1}}, this phase is specified by a dot-dashed curve. 
The above is a conventionally accepted fact. 
In the region $x>1$, there exist two phases and the energy of the 
superconducting phase is smaller than that of the normal conducting phase. 
From this reason, we chase the system under investigation along the 
solid to the broken line, the connection of which is not smooth. 
In this sense, conventionally 
we call this situation the sharp phase transition. 

Our next task is to investigate the system under the condition 
$\hbar=$finite. 
For this purpose, it may be enough to solve the quadratic equation 
with respect to $x$, (\ref{4-5}). 
In order to see the whole feature of the solution, for a moment, 
we treat the variables $\rho$ and $x$ in the region 
\begin{equation}\label{4-16}
-1 \leq \rho \leq 1 \ , \qquad -\infty < x < \infty \ . 
\end{equation}
Two solutions of Eq.(\ref{4-5}) for the case $\kbar\neq 0$ and 1 are 
given exactly in simple forms: 
\begin{subequations}\label{4-17}
\beqn
& &x=\frac{1}{\sqrt{1-\rho^2}}\cdot 
\frac{\rho^3-\kbar}{\rho^2\left(\rho-\frac{\kbar}{2}\right)} \ , 
\label{4-17a}\\
& &x=-\frac{1}{\sqrt{1-\rho^2}} \ . 
\label{4-17b}
\eeqn
\esub
The value of $\rho$ at the cross point of the solutions (\ref{4-17a}) 
and (\ref{4-17b}), which we denote $\rho_P$, is given by 
\beqn\label{4-18}
& &\rho_P
=\left(\frac{\kbar}{2}\right)^{\frac{1}{3}}
\biggl[
\left(1+\frac{1}{2}\cdot \frac{\frac{1}{54}\left(\frac{\kbar}{2}\right)^2}
{1+\sqrt{1+\frac{1}{54}\left(\frac{\kbar}{2}\right)^2}}\right)^{\frac{2}{3}}
\nonumber\\
& &\qquad\qquad\qquad\quad
+\!
\left(\frac{1}{2}\cdot \frac{\frac{1}{54}\left(\frac{\kbar}{2}\right)^2}
{1+\sqrt{1+\frac{1}{54}\left(\frac{\kbar}{2}\right)^2}}\right)^{\frac{2}{3}}
\!\!+\!\frac{1}{6}\left(\frac{\kbar}{2}\right)^{\frac{2}{3}}\biggl] \ .
\eeqn
\begin{figure}[t]
\begin{center}
\includegraphics[height=9.0cm]{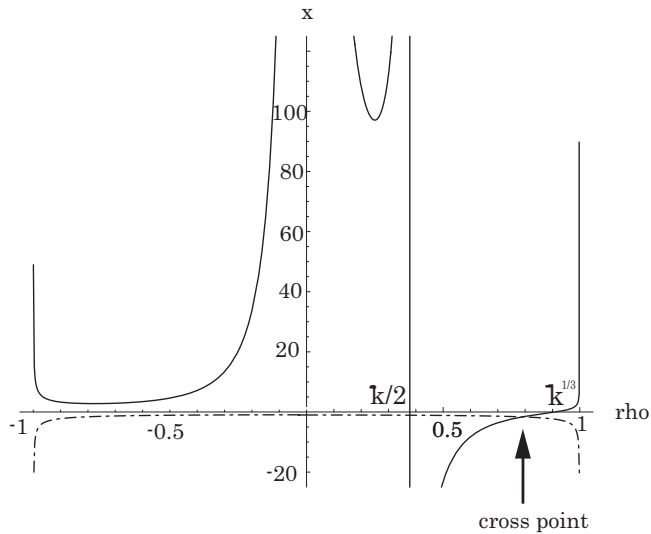}
\caption{The variable $x$ is depicted as a function of 
$\rho$ with $\hbar\neq 0\ (\kbar=0.75)$.
}
\label{fig:4-2}
\end{center}
\end{figure}
Figure {\ref{fig:4-2}} shows the behavior of $x$ as a function of $\rho$. 
The solid curves show the behaviors of $x$ given in the 
relation (\ref{4-17a}) in the region 
$(0\leq \rho \leq 1\ , \ 0\leq x < \infty)$ and 
in the other region, respectively. 
The dot-dashed curve shows the behavior of the solution 
(\ref{4-17b}). 
We can see that between Figs.{\ref{fig:4-1}} and {\ref{fig:4-2}}, 
there exists quite big difference. 
In \S 6, we will discuss this difference. At first sight, 
Figs.{\ref{fig:4-1}} and {\ref{fig:4-2}} tell us that, under the 
limit $\hbar\ (\kbar)\ \rightarrow 0$, the solid curve in the region 
$\kbar^{1/3}\leq \rho \leq 1$ of Fig.{\ref{fig:4-2}} approaches to the 
solid line and the broken curve in Fig.{\ref{fig:4-1}} and 
the solid curve in the region 
$0\leq \rho \leq \kbar^{1/3}$ is compressed and becomes the dotted line in 
Fig.{\ref{fig:4-1}}. 
Through the discussion in \S 6, the above view will be answered in the 
negative.

\section{Approximate solution for the relation determining the 
energy minimum point}

As was shown in Eqs.(\ref{4-17a}) and (\ref{4-17b}), we can express 
$x$ in terms of $\rho$. 
However, in order to understand the behavior of the system, 
for example, the change of the energy, in relation to the interaction 
strength, we must express $\rho$ as a function of $x$. 
Numerically, it is possible and easy. 
But, analytical expression may be also helpful for the understanding of the 
system. 
In the present case, the exact form cannot be almost expected and we will 
try to the approximate form.

Let us investigate the case of the smooth curve in the region 
$\kbar/2 \leq \rho \leq 1$ in Fig.{\ref{fig:4-2}}. 
First, we note Eq.(\ref{4-17a}). 
This equation can be rewritten in the form 
\begin{equation}\label{5-1}
x=\frac{\rho-\kbar^{1/3}}{(1-\rho)(\rho-\kbar/2)}\cdot 
\sqrt{\frac{1-\rho}{1+\rho}}\cdot\frac{\rho^2+\kbar^{1/3}\rho+\kbar^{2/3}}
{\rho^2} \ . 
\end{equation}
Then, we define the quantity $y$ in the form 
\begin{equation}\label{5-2}
y=\sqrt{\frac{1+\rho}{1-\rho}}\cdot\frac{\rho^2}{\rho^2+\kbar^{1/3}\rho
+\kbar^{2/3}}\cdot x \ . 
\end{equation}
With the quantity $y$, the relation (\ref{5-1}) is formally reduced to 
the quadratic equation for $\rho$: 
\begin{equation}\label{5-3}
y=f(\rho) \ , \qquad
f(\rho)=\frac{\rho-\kbar^{1/3}}{(1-\rho)(\rho-\kbar/2)} \ . 
\end{equation}
Formally, from the relation (\ref{5-3}), we have 
\beqn\label{5-4}
& &\rho=F(y) \ , \nonumber\\
& &F(y)=\frac{\kbar y-2\kbar^{1/3}}
{\left(1+\frac{\kbar}{2}\right)y-1-
\sqrt{\left(1-\frac{\kbar}{2}\right)^2y^2-2\left(1+\frac{\kbar}{2}-2\kbar^{1/3}
\right)y+1}} \ . 
\eeqn
If $x\rightarrow -\infty$, 0 and $+\infty$, i.e., 
$y\rightarrow -\infty$, 0 and $+\infty$, we have the relations 
$F(-\infty)=\kbar/2$, $\kbar^{1/3}$ and 1. 
Therefore, at the three characteristic points, we are able to obtain the exact 
results. 
From this reason, we do not adopt another branch for the solutions of the 
quadratic equation.

Next, we consider the case of the points except $x\rightarrow -\infty$, 
0 and $+\infty$. 
First, we define a series $\{ \rho^{(0)}, \rho^{(1)},\rho^{(2)},\cdots\}$ 
obeying the following recursion relation: 
\beqn\label{5-5}
& &\rho^{(k)}=F(y^{(k-1)}) \ , \qquad y^{(k-1)}=f(\rho^{(k-1)}) \ , 
\nonumber\\
& &k=1,2,3,\cdots . 
\eeqn
Here, we note that the function $f(y)$ is monotone-increasing. 
Then, if there exist the limiting value 
$\lim_{k\rightarrow \infty}\rho^{(k)}=\rho$, 
$\rho$ is determined as a function of $\rho$ by the condition 
\begin{equation}\label{5-6}
\rho=F(y) \ , \qquad y=f(\rho) \ .
\end{equation}
The relation (\ref{5-6}) is nothing but the set of the relations 
(\ref{5-3}) and (\ref{5-4}). 
Under appropriate choice for the initial value $\rho^{(0)}$, iteratively, 
we obtain $\rho^{(k)}$ from $k=1$ and after infinite iteration, we may 
obtain the exact solution which do not depend on $\rho^{(0)}$. 
If we stop the iteration half way, the result is an approximated 
one which depends on $\rho^{(0)}$. 
As a possible choice of the initial condition, we adopt the value 
of $\rho$ satisfying the form (\ref{4-17a}) at $x=1$, which will be denoted as 
$\rho_c$: 
\begin{equation}\label{5-7}
1=\frac{1}{\sqrt{1-\rho_c^2}}
\frac{\rho_c^3-\kbar}{\rho_c^2(\rho_c-\kbar/2)} \ .
\end{equation}
The relation (\ref{5-7}) leads us to 
\begin{equation}\label{5-8}
\kbar=\frac{\rho_c^5}{(1+\sqrt{1-\rho_c^2})
\left(1-\frac{\rho_c^2}{2}\sqrt{1-\rho_c^2}\right)} \ .
\end{equation}
Since the right-hand side of the relation (\ref{5-8}) is 
increasing, and then, $\kbar$ and $\rho_c$ are in the relation of 
one-to-one correspondence. 
Therefore, instead of $\kbar$, we can use $\rho_c$ defined in the 
relation (\ref{5-8}). 
Therefore, the cases $x=-\infty$, 0, 1 and $+\infty$ do not 
depend on the iteration.

\section{Examination in the case $\kbar\rightarrow 0$ and 1 and its related 
phase transition}

In \S 4, we showed the behavior of $x$ with respect to $\rho$ in the 
case where $\kbar$ does not approach to 0 and 1. 
In this section, we discuss the case where $\kbar$ approaches 
to 0 and 1. 
First, we treat the case $\kbar\rightarrow 0$. 
In the case $\rho<\rho_c$, we have 
\begin{equation}\label{6-1}
x=\frac{\rho^3-\kbar}{\sqrt{1-\rho^2}\rho^2\left(\rho-\frac{\kbar}{2}\right)}
\longrightarrow \frac{1}{\sqrt{1-\rho^2}} \ , \qquad (\kbar\rightarrow 0) \ . 
\end{equation}
Further, in the case $\rho=\rho_c$, we have 
\begin{eqnarray*}
& &1
=\frac{\rho_c^3-\kbar}{\sqrt{1-\rho_c^2}\rho_c^2
\left(\rho_c-\frac{\kbar}{2}\right)}
\longrightarrow \frac{1}{\sqrt{1-\rho_c^2}}\ , \qquad (\kbar\rightarrow 0) \ , 
\end{eqnarray*}
namely, 
\begin{equation}\label{6-2}
\rho_c \longrightarrow 0 \ . \qquad (\kbar \rightarrow 0)
\end{equation}
In the case where $\rho$ is a little bit larger than $\kbar^{1/3}$, 
we can put $\rho=(\alpha\kbar)^{1/3}$ $(\alpha>1)$. 
Then, we have 
\bsub\label{6-3}
\beqn
x&=&\frac{\alpha\kbar-\kbar}{\sqrt{1-(\alpha\kbar)^{2/3}}
(\alpha\kbar)^{2/3}\left((\alpha\kbar)^{1/3}-\frac{\kbar}{2}\right)}
\nonumber\\
&=&\frac{\alpha-1}{\sqrt{1-(\alpha\kbar)^{2/3}}\left(\alpha-\frac{1}{2}
\alpha^{2/3}\kbar^{2/3}\right)}
\longrightarrow 1-\frac{1}{\alpha}\ , \quad (\kbar\rightarrow 0)\ . 
\label{6-3a}
\eeqn
The above relation leads us to 
\begin{equation}\label{6-3b}
\rho \longrightarrow 0 \ , \qquad x\longrightarrow 1-\frac{1}{\alpha}\ , 
\qquad \left(0<1-\frac{1}{\alpha}<1\right)\ , \quad
(\kbar\rightarrow 0) \ .
\end{equation}
\esub
\begin{figure}[t]
\begin{center}
\includegraphics[height=3.8cm]{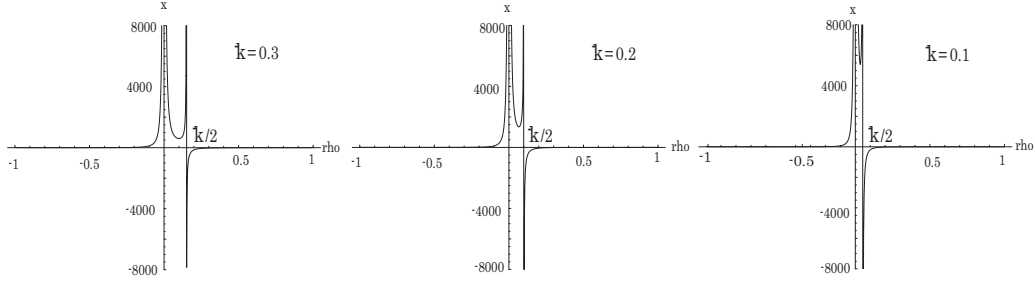}
\caption{The variable $x$ is depicted as a function of 
$\rho$ with $\kbar\rightarrow 0$ $(\kbar=0.3,\ 0.2,\ 0.1)$.
}
\label{fig:6-0a}
\end{center}
\end{figure}

\noindent
The relation (\ref{6-3b}) tells that under the limit $\kbar\rightarrow 0$, 
$\rho\rightarrow 0$ and $x$ takes arbitrary value between 0 and 1. 
From the above three cases, we see that $x$ in the region 
$\kbar^{1/3}<\rho <1$ in Fig.{\ref{fig:4-2}} is reduced to the solid line 
and the dot-dashed curve in Fig.\ref{fig:4-1}. 
Next, we treat the case where $\rho$ is a little bit smaller than 
$\kbar^{1/3}$. 
In this case, we can set up $\rho=(\kbar/\alpha)^{1/3}\ (\alpha>1)$. 
Then, in the same process, we have 
\bsub\label{6-4}
\begin{equation}\label{6-4a}
x\longrightarrow 1-\alpha\ , \qquad (\kbar\rightarrow 0) \ . 
\end{equation}
The above means the following situation: 
\begin{equation}\label{6-4b}
\rho\longrightarrow 0\ , \qquad x\longrightarrow 1-\alpha \ , 
\qquad (1-\alpha<0) \ . \qquad (\kbar\rightarrow 0)
\end{equation}
\esub
The above situation can be interpreted as that under $\kbar\rightarrow 0$, 
$\rho\rightarrow 0$ and $x$ takes negative arbitrary value and 
in Fig.{\ref{fig:4-1}}, the straight line between 0 and 1 continues 
to the negative region of $x$. 
In the case where $\rho$ is in the region $0<\rho<\kbar/2$, 
we can put $\rho=\alpha(\kbar/2),\ (0<\alpha<1)$. 
Then, we have 
\beqn\label{6-5}
x&=&\frac{(\alpha\kbar/2)^3-\kbar}{\sqrt{1-(\alpha\kbar/2)^2}
(\alpha\kbar/2)^2(\alpha\kbar/2-\kbar/2)} \nonumber\\
&=&\frac{8/\kbar^2-\alpha^3}{\sqrt{1-\alpha^2\kbar^2/4} \ \alpha(1-\alpha)}
\longrightarrow +\infty \ . 
\qquad (\kbar\rightarrow 0)
\eeqn
This situation is very interesting. 
The solid curve in the region $0<\rho<\kbar/2$ in Fig.\ref{fig:4-2} 
disappears to $+\infty$. 
Therefore, the dotted line in Fig.\ref{fig:4-1} does not correspond to 
this curve. 
Next, in the case where $\rho$ is a little bit smaller than 0, we can set up 
$\rho=-(\alpha\kbar)^{1/3}\ (\alpha>0)$. 
Then, $x$ can be expressed in the form 
\bsub\label{6-6}
\beqn\label{6-6a}
x&=&\frac{\alpha\kbar+\kbar}{\sqrt{1-(\alpha\kbar)^{2/3}}
\alpha^{2/3}\kbar^{2/3}\left(\alpha^{1/3}\kbar^{1/3}+\frac{\kbar}{2}\right)}
\nonumber\\
&=&\frac{1}{\sqrt{1-(\alpha\kbar)^{2/3}}}\cdot
\frac{\alpha+1}{\alpha+\frac{1}{2}
\alpha^{2/3}\kbar^{2/3}}
\longrightarrow 1+\frac{1}{\alpha}\ , \qquad (\kbar\rightarrow 0)\ . 
\eeqn
The above means 
\begin{equation}\label{6-6b}
\rho\longrightarrow 0\ , \qquad 
x\longrightarrow 1+\frac{1}{\alpha}\ , \qquad 
\left(1+\frac{1}{\alpha}>1\right)\ . \qquad (\kbar\rightarrow 0)
\end{equation}
\esub
In this case, under the limit $\kbar\rightarrow 0$, $\rho \rightarrow 0$ and 
$x$ is positive arbitrary and this situation corresponds to the 
dotted line starting at $x=1$ in Fig.\ref{fig:4-1}. 
However, this situation comes from the negative region of $\rho$, which 
is physically unaccepted. 
Further, in the case $\rho \ll -(\alpha\kbar)^{1/3}$, 
$x\rightarrow 1/\sqrt{1-\rho^2}\ (\kbar\rightarrow 0)$.

\begin{figure}[t]
\begin{center}
\includegraphics[height=3.8cm]{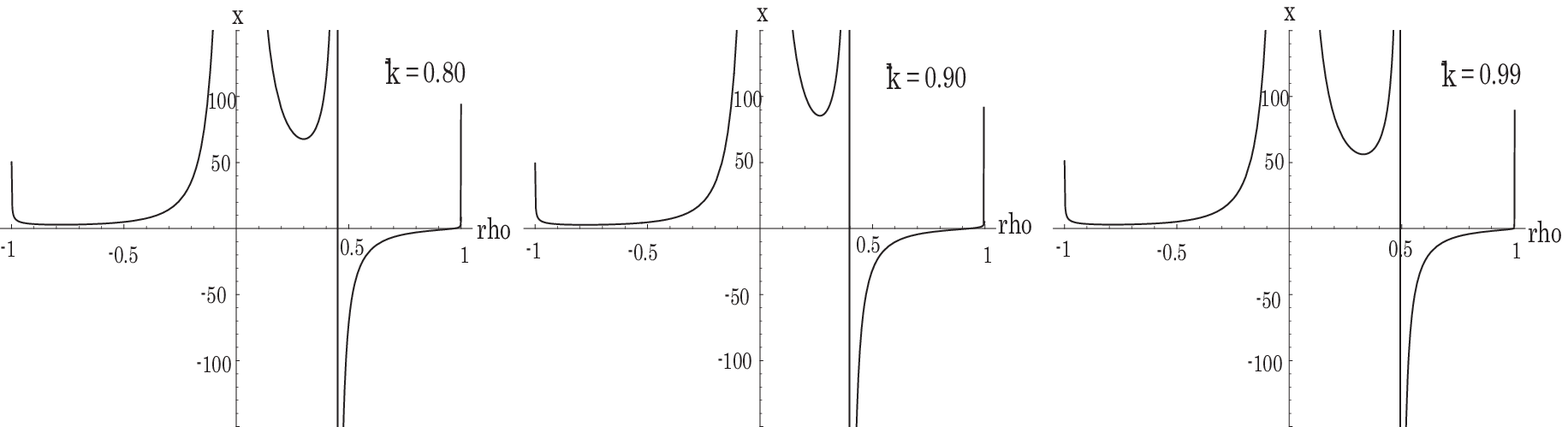}
\caption{The variable $x$ is depicted as a function of 
$\rho$ with $\kbar\rightarrow 1$ $(\kbar=0.8,\ 0.9,\ 0.99)$.
}
\label{fig:6-0b}
\end{center}
\end{figure}

Next, we treat the case $\kbar\rightarrow 1$. 
In this case, we divide two situations. 
One is the case $\kbar^{1/3}<\rho$. 
If we put $\kbar=1-\delta\kbar$ and $\rho=1-\delta\rho$, we have 
$\delta\rho<\delta\kbar/3$. 
Then, we put $\delta\rho=\alpha\delta\kbar/3$ $(0<\alpha<1)$.
Under the condition that $\delta\kbar\ (\delta\rho)$ is sufficiently small, 
$x$ can be expressed as 
\begin{equation}\label{6-7}
x=\frac{(1-\alpha)\delta\kbar}{\frac{2}{3}\alpha\delta\kbar\times
\frac{1}{2}}
\longrightarrow 3\left(\frac{1}{\alpha}-1\right)\ , 
\qquad \left(\frac{1}{\alpha}-1>0\right)\ . 
\qquad (\kbar\rightarrow 0)
\end{equation}
In this case, under the limit $\kbar\rightarrow 1$, $\rho\rightarrow 1$ and 
$x$ is positive arbitrary. 
The above means that the solid curve between $\rho=\kbar^{1/3}$ and 1 
becomes the straight line on $\rho=1$. 
On the other hand, for the region $\kbar/2\leq \rho \leq \kbar^{1/3}$, 
the limit $\kbar\rightarrow 1$ gives 
\begin{equation}\label{6-8}
x\longrightarrow -\sqrt{\frac{1-\rho}{1+\rho}}\frac{1+\rho+\rho^2}
{\rho^2(\rho-1/2)}\ . 
\end{equation}
The above shows that the solid curve between $\rho=\kbar/2$ and 
$\kbar^{1/3}$ in Fig.\ref{fig:4-2} becomes the curve showing by the relation 
(\ref{6-8}). 
The other cases do not have characteristic change. 
However, as was already mentioned, $\kbar\rightarrow 1$ corresponds 
to $L\rightarrow 0$. 
In the system in which the seniority is zero, $L$ starts from $L=1/2$, 
and then, the present case is not so interesting. 
But, in the case where the seniority is not zero, it may be interesting. 


\begin{figure}[t]
\begin{center}
\includegraphics[height=4.0cm]{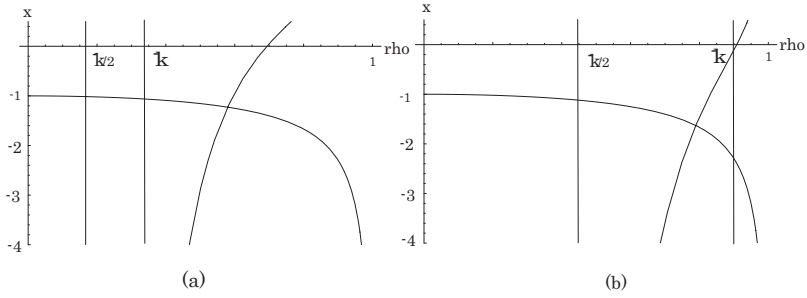}
\caption{The cross points are shown in the case (a) $\kbar=0.33$ 
and (b) $\kbar=0.90$.
}
\label{fig:6-1a}
\end{center}
\end{figure}

\begin{figure}[b]
\begin{center}
\caption{The three types of non-smoothness and discontinuity for 
the crossing point $\rho_P$
are illustrated in the case $\kbar<\rho_P$.
}
\label{fig:6-2}
\end{center}
\end{figure}

On the basis of the above-shown results, let us consider the phase transition 
in our model. 
In the present case, physically interesting feature can be seen in the solid 
curve between $\rho=\kbar^{1/3}$ and 1 in Fig.\ref{fig:4-2}. 
We cannot find any non-smoothness or discontinuity. 
Of course, under the limit $\kbar\rightarrow 0$, the result appears 
as the phase transition which is conventionally accepted. 
Therefore, in the finite $\hbar$ and $\lambda$, we should not expect the phase 
transition. 
Further, we cannot find any curve which leads us to the dotted line in 
Fig.\ref{fig:4-1} from the side of the physically acceptable region. 
Therefore, the dotted line in Fig.\ref{fig:4-1} should be negative 
understood. 
From the above examination, we can learn that the limiting process 
$\hbar\rightarrow 0$ must be carefully performed. 
All do not follow this process. 

\begin{figure}[b]
\begin{center}
\caption{A type of non-smoothness and discontinuity for 
the crossing point $\rho_P$ is illustrated in the case $\kbar > \rho_P$.
}
\label{fig:6-3}
\end{center}
\end{figure}

Let our model accepted in the region $x<0$ (the repulsive 
pairing interaction). 
In Fig.\ref{fig:4-2}, we see that there exists a cross point of two 
different curves. 
Noting the relation (\ref{4-11}) $(\kbar\leq \rho\leq 1)$, 
we examine characteristic feature of the crossing. 
There exist two types of the crossing which show in 
Fig.{\ref{fig:6-1a}}(a) and \ref{fig:6-1a}(b). 
Both figures are characterized by $\rho_P>\kbar$ and $\kbar/2<\rho_P<\kbar$, 
respectively. Here, $\rho_P$ is shown in the relation (\ref{4-18}). 
From Fig.\ref{fig:6-1a}, we learn that there exist three types of 
non-smoothness and discontinuity, which are illustrated by 
Fig.\ref{fig:6-2}. 
Following the change of the interaction strength, the system moves 
along the solid curve. 
At the present, we do not judge which should be chose in the three cases. 
In the case of Fig.\ref{fig:6-1a}(b), we can draw the picture shown in 
Fig.\ref{fig:6-3}. 
Discontinuity appears in the interaction strength. 
Thus, we can conclude that in our present model, the phase transition 
appears in the case of the negative interaction strength. 
However, at the present stage, we cannot judge if the above phase 
transition has its realistic meaning.


\section{Discussion}

\begin{figure}[b]
\begin{center}
\caption{The ground state energy normalized by $\epsilon$ 
is depicted as a function of the interaction 
strength $x=4G\lambda/\epsilon$ with $\kbar=1/3$ (solid curve) 
together with $E_0$ and $E_1$.
}
\label{fig:7-1}
\end{center}
\end{figure}

Finally, in this section, we discuss some characteristic points of the 
energy $E$. 
The energy $E$ is obtained by substituting the solution of Eq.(\ref{4-1}) 
into $H$ shown in the relation (\ref{3-22}): 
\begin{equation}\label{7-1}
E=-\epsilon\cdot 2\lambda 
\left[\left(1-\frac{\kbar}{\rho}\right)\sqrt{1-\rho^2}
+\frac{x}{2}\rho(\rho-\kbar)\right] \ .
\end{equation}
In Fig.{\ref{fig:7-1}}, we show numerical results. 
In the case $x\rightarrow \infty$ (two levels are degenerate), 
we have $\rho=1$ and $E$ is reduced to 
\begin{equation}\label{7-2}
E_1=-G\cdot 2L(2L+\hbar)=-\hbar^2 G\Omega(\Omega+1) \ . 
\end{equation}
The form (\ref{7-2}) is the same as that shown exactly. 
On the other hand, in the case $x=0$ (free fermions), we have 
$\rho=\kbar^{1/3}$ and $E$ is written as 
\beqn\label{7-3}
E_0&=&-\epsilon\cdot 2\lambda\left(1-\kbar^{\frac{2}{3}}\right)^{\frac{3}{2}}
\nonumber\\
&=&-\hbar\epsilon\left(\frac{2L+\hbar}{\hbar}\right)\left(
1-\left(\frac{\hbar}{2L+\hbar}\right)^{\frac{2}{3}}\right)^{\frac{3}{2}}
\nonumber\\
&=&-\frac{\hbar\epsilon}{2}\cdot 2\left[(\Omega+1)^{\frac{2}{3}}-1
\right]^{\frac{3}{2}} \ . 
\eeqn
If $\Omega$ is large, the leading term of $E$ is given by 
\begin{equation}\label{7-4}
E \approx -\frac{\hbar\epsilon}{2}\cdot 2\Omega\ 
\left(=-\frac{\hbar\epsilon}{2}N \right) \ . 
\end{equation}
The above form tells us that the lower level is fully occupied. 
Therefore, we can conclude that in the case of large $\Omega$, our result is 
reduced to the exact one and for the problem of $\kbar\rightarrow 0$, 
it may be favorable.

\section{Summary}

In this paper, we analyzed the aspect of phase change in the two-level 
pairing model governed by the $su(2)\otimes su(2)$-algebra. 
This model can be described in terms of four kinds of boson operators 
by means of the Schwinger boson representation of the $su(2)$-algebras. 
We used the variational approach with the mixed-mode 
coherent state introduced in Ref.\citen{A}. This state is constructed, 
as is similar to the usual coherent state,  
by using the raising operators of the $su(2)$- and the $su(1,1)$-algebras 
whose generators are defined by the Schwinger bosons used in this model. 

In the classical limit $\hbar \rightarrow 0$, 
in which the Planck constant is taken to be 0 at the first time, 
if the force strength is smaller than a certain critical value, 
the trivial phase, which is called the non-condensed phase, is realized. 
At the critical point with respect to the force strength, a new branch 
appears and this branch reveals the existence of the so-called 
condensed phase. However, the non-condensed phase still exists as a result of 
the trivial solution if the force strength is larger than the critical value. 
This situation is well known in the usual Hartree-Bogoliubov approximation 
and the sharp phase transition is realized in this treatment. 

However, this behavior is not true because the sharp phase transition 
is not realized in this many-body system and this fact can easily be verified 
in this exactly solvable algebraic model. 
Figures \ref{fig:4-2} shows the aspect of phase structure. Furthermore, 
in Fig.\ref{fig:6-0a}, 
the aspect of the phase change in the semi-classical limit 
is shown clearly. 
It is seen in Fig.\ref{fig:4-2} 
that the two branches exist in the physical region in 
$0 < \rho \leq 1$, where 0 is included when $\hbar \rightarrow 0$. 
If $\hbar$ approaches to 0, one branch, which is located in $0<\rho<\kbar/2$, 
goes to infinity of $x$ 
and this branch disappears in the limit $\hbar\rightarrow 0$. 
Another branch, which is located in $\kbar/2 < \rho \leq 1$ and $x\geq 0$, 
is reduced to the physical branch in Fig.{\ref{fig:4-1}}, 
which is represented by 
solid and dash-dotted curves. 
However, the branch of ``the non-condensed phase'' in the classical limit, 
which is 
represented by the dotted line in Fig.1 in $x>1$, appears from the 
unphysical region in $\rho<0$ in Fig.\ref{fig:4-2}. 
So, this solution is not a physical 
solution originally. Thus, this ``non-condensed phase" is not realized 
in the physical system. 
Therefore, the sharp phase transition does not occur if $\hbar$ is taken 
as finite value. 

In our model, there exist a cross point of two branches represented by 
two curves in Fig.\ref{fig:4-2} in the negative region of $x$. 
Because the variable $x$ 
is proportional to the force strength $G$, this cross point is 
realized in the case of the ``repulsive pairing". 
Thus, the sharp phase transition may be realized in the case of the negative 
interaction strength.

\section*{Acknowledgements} 
This work started when two of the authors (Y. T. and M. Y.) 
stayed at Coimbra in September 2005 and was completed when 
Y. T. stayed at Coimbra in September 2006. They would like to 
express their sincere thanks to Professor Jo\~ao da Provid\^encia, a 
co-author of this paper, for his invitation and warm hospitality. 
One of the authors (Y. T.) 
is partially supported by a Grant-in-Aid for Scientific Research 
(No.15740156 and No.18540278) 
from the Ministry of Education, Culture, Sports, Science and 
Technology of Japan.


\appendix

\section{A method for solving the relation (\ref{5-5})}

Let an infinite series $\{y^{(0)},y^{(1)},y^{(2)}, \cdots\}$ obey 
the following condition characterized by a function $f(x,y)$:
\begin{equation}\label{a-1}
y^{(k)}=\left(f(x,y^{(k-1)})\right)^\nu\ . 
\qquad (k=1,2,3,\cdots)
\end{equation}
Here, $\nu$ denotes real parameter and $x$ plays a role of real variable 
$(0\leq x <\infty)$. 
Further, we regard $f(x,y)$ as a monotone-increasing or -decreasing function.
If there exists the limiting value 
$\lim_{k\rightarrow \infty}y^{(k)}\ (=y)$, $y$ is determined as a function 
of $x$ by the condition 
\begin{equation}\label{a-2}
y=\left(f(x,y)\right)^\nu \ .
\end{equation}
The above means that the variable $y$ governed by the relation (\ref{a-2}) 
is expressed as a function of $x$ by calculating the limiting value 
$\lim_{k\rightarrow \infty}y^{(k)}$ of the series\break 
$\{y^{(0)},y^{(1)},y^{(2)}, \cdots\}$ obeying the condition (\ref{a-1}). 
On the basis of the above idea, in this Appendix, we develop a 
systematic method for obtaining an approximate expression of $y$ 
as a function of $x$. 
Of course, $x$ and $y$ obey the relation (\ref{a-2}), and hereafter, 
we treat the following case: 
Through the relation (\ref{a-2}), we are able to obtain the values of 
$y$ at $x=0$ and at $x\rightarrow \infty$. 
These two values are denoted as $y_0$ and $y_\infty$, respectively: 
\begin{equation}\label{a-3}
y_0=\left(f(0,y_0)\right)^\nu \ , \qquad
y_\infty=\left(f(\infty,y_\infty)\right)^\nu \ . 
\end{equation}
Hereafter, we denote any quantity $X$ at $x=0$ and $x\rightarrow \infty$ as 
$X_0$ and $X_\infty$, respectively. 
The derivative $(dy/dx)_0$ is given as 
\begin{equation}\label{a-4}
\left(\frac{dy}{dx}\right)_0
=\frac{\left(\frac{\partial f}{\partial x}\right)_0}
{\frac{1}{\nu y_0^{\nu-1}}-\left(\frac{\partial f}{\partial y}\right)_0} \ .
\end{equation}
The quantities $(\partial f/\partial x)_0$ and $(\partial f/\partial y)_0$ 
are obtained by putting $x=0$ and $y=y_0$ for $\partial f/\partial x$ 
and $\partial f/\partial y$ as functions of $x$ and $y$, respectively. 

Let us set up the following form for $y^{(0)}$: 
\begin{equation}\label{a-5}
y^{(0)}=\left(\frac{y_\infty^{\frac{1}{\nu}}x+y_0^{\frac{1}{\nu}}\alpha}
{x+\alpha}\right)^\nu \ .
\end{equation}
Independently from $\alpha$, we have 
\begin{equation}\label{a-6}
y_0^{(0)}=y_0 \ , \qquad y_\infty^{(0)}=y_\infty \ . 
\end{equation}
We choose $\alpha$ so as to make $y^{(0)}$ satisfy the relation (\ref{a-4}):
\begin{equation}\label{a-7}
\alpha=\nu y_0\left[\left(\frac{y_\infty}{y_0}\right)^{\frac{1}{\nu}}
-1\right]\frac{\frac{1}{\nu y_0^{\nu-1}}-\left(\frac{\partial f}{\partial y}
\right)_0}{\left(\frac{\partial f}{\partial x}\right)_0} \ . 
\end{equation}
Therefore, naturally, we have 
\begin{equation}\label{a-8}
\left(\frac{d y^{(0)}}{dx}\right)_0=\left(\frac{d y}{dx}\right)_0 \ . 
\end{equation}
Through the relation (\ref{a-1}), we obtain $y^{(1)}$ from $y^{(0)}$ given 
by the form (\ref{a-5}). 
By performing the above process successively, we can calculate 
$y^{(k)}$ for any $k$. 
If $k$ is larger, the accuracy of the approximation may be better. 
We can prove the following form: 
\beqn\label{a-9}
& &y_0^{(k)}=y_0^{(0)}=y_0 \ , \qquad
y_\infty^{(k)}=y_\infty^{(0)}=y_\infty \ , \nonumber\\
& &\left(\frac{dy^{(k)}}{dx}\right)_0=\left(\frac{dy^{(0)}}{dx}\right)_0=
\left(\frac{dy}{dx}\right)_0 \ . \quad (k=2,3,4,\cdots)
\eeqn
Therefore, we can conclude that, for any $k$, the behaviors of $y^{(k)}$ 
near $x=0$ and $x\rightarrow \infty$ do not change from those for $k=0$.


\begin{thebibliography}{99}
\bibitem{3}
A. Kuriyama, J. da Provid\^encia, C. Provid\^encia, Y. Tsue and M. Yamamura, 
Prog. Theor. Phys. {\bf 95} (1996), 339.
\bibitem{1}
H. J. Lipkin, N. Meshkov and A. J. Glick, Nucl. Phys. {\bf 62} (1965), 188. 
\bibitem{2}
A. K. Kerman, Ann. of Phys. {\bf 12} (1961), 300. 
\bibitem{4}
Y. Tsue, C. Provid\^encia, J. da Provid\^encia and M. Yamamura, 
submitted to Prog. Theor. Phys.
\bibitem{A}
A. Kuriyama, J. da Provid\^encia and M. Yamamura, 
Prog. Theor. Phys. {\bf 103} (2000), 305.
\end{thebibliography}
\end{document}